\documentclass[a4pa per,11pt]{article} 
\usepackage[english]{babel}           
\usepackage[utf8]{inputenc}           

\usepackage{graphicx}    
\usepackage{color}       
\usepackage{anysize}     
\usepackage{multicol}    
\usepackage{bm}          %
\usepackage{subcaption}
\usepackage{eurosym}     
\usepackage{amsthm}      %
\usepackage{gensymb}
\usepackage{ amssymb }
\usepackage{float}
\usepackage{lineno}      %
\usepackage{multirow}
\usepackage{dsfont} 
\usepackage{hyperref}
\usepackage{cite} 
\usepackage{mciteplus}
\usepackage{ifthen}
\usepackage{unravel}

\usepackage[nottoc]{tocbibind}

\marginsize{2.5cm}{2.5cm}{1.5cm}{1.5cm} %
\begin{document}

\begin{titlepage}
	
\vfill

\begin{center}
	
	{\centering\LARGE\bf Calibration of the momentum scale of a particle physics detector using the Armenteros-Podolanski plot}
	
	\vspace{2cm}
	Pablo Baladr\'on Rodr\'iguez$^1$, Veronika Chobanova $^1$, Xabier Cid Vidal$^1$, Vladimir Gligorov$^2$, Miriam Lucio Mart\'inez$^3$, Jovan Markov$^4$,  Diego Mart\'inez Santos$^1$, M\'aximo Pl\'o Casas\'us$^1$
	
	{\it
$^1$Instituto Galego de F\'isica de Altas Enerx\'ias (IGFAE), Universidade de Santiago de Compostela, Santiago, Spain \\
$^2$LPNHE, Sorbonne Universit\'e, Paris Diderot Sorbonne Paris Cit\'e, CNRS/IN2P3, Paris, France\\
$^3$Nikhef National Institute for Subatomic Physics, Amsterdam, The Netherlands\\
$^4$Weizmann Institute, Department of Particle Physics and Astrophysics, Rehovot 7610001, Israel
}
\end{center}
\vspace{2cm}
\begin{abstract}

A method for calibrating the momentum scale in a particle physics detector is described. The method relies on the determination of the masses of the final state particles in two-body decays of neutral particles, which can then be used to obtain corrections in the momentum scale. A modified version of the Armenteros-Podolanski plot and the $K_S^0 \to \pi^+ \pi^-$ decay is used as a proof of principle for this method.  
A precision at the $10^{-6}-10^{-8}$ level is achieved in simplified simulations.

\end{abstract}

\end{titlepage}

\newpage
\tableofcontents
\newpage

\section{Introduction}

The Armenteros-Podolanski plot~\cite{armenteros} is a representation of the transverse momentum of a two-body decay versus the asymmetry of the longitudinal momentum of the decay products. In this space two-body decays appear as semi-ellipses, whose parameters provide information on the masses of the parent and the child particles.

This representation was proposed in 1954 by R. Armenteros and J. Podolanski as a method of analysis of the dynamics of neutral particles decaying to two bodies ($V^0$ particles). It was used to separate $K_S^0 \rightarrow \pi^+ \pi^- $ from $\Lambda^0 \rightarrow p \pi^-$ decays without any assumptions about the masses of the final state decay products. Conversely, by fitting the ellipses observed
in data, one can retrieve the masses of the parent and the child particles. Deviations from known values then
give useful information about detector and reconstruction effects, in particular biases in the momentum scale, which can be caused by, e.g., imperfect knowledge of the magnetic field of a detector.

At present, the calibration of the momentum scale of LHC detectors uses the fit of the invariant mass of known resonances. With this method, e.g., the LHCb experiment achieves a precision of $10^{-4}$~\cite{detector_performance,Aaij:2011ep} using several resonances, mostly $J/\psi\rightarrow\mu^+\mu^-$ decays. The precision of this method is ultimately limited by the knowledge of such masses, and understanding of the dependency of those uncertainties with other inputs. Our objective is to demonstrate that the Armenteros-Podolanski plot allows a significantly more precise momentum scale calibration than using only parent particles, as long as child particle masses are better known than parent particle masses (though the later are also expected to improve with time). In this paper, a proof of principle is shown, using $V^0$ decays generated with Pythia 8 ~\cite{pythia} at Run~2 LHC energies (13 TeV), and artificially introducing a set of different biases in the momentum scale. We do not include momentum resolution effects in this study, although we provide an estimate on the expected uncertainty that would arise from it.

The key advantage of our proposed method is that it uses the precisely known masses of the final state particles of the decay (e.g, pions, protons, muons...) to calibrate the momentum scale of the detector, instead of the less precisely measured masses of the heavier parent particles. For simplicity, in this work we use the pion mass~\cite{pion_mass} in the $K_S^0\to \pi^{+}\pi^{-}$ decay, although a better precision could be achieved using the proton mass in the $\Lambda^0 \rightarrow	p \pi$ decay \cite{PDG2018}. However, fitting the $\Lambda^0$ ellipses is more challenging since the two decay products have different masses, which results in a higher number of fit parameters.

\section{The Armenteros-Podolanski space}
The Armenteros-Podolanski space \cite{armenteros} is defined for a two-body decay ($M \rightarrow m_1 m_2$) as ($p_T^{\phantom{2}} - \alpha$), where $\alpha$ is the longitudinal momentum asymmetry:

$$\alpha = \frac{p_{L1}^{\phantom{2}}-p_{L2}^{\phantom{2}}}{p_{L1}^{\phantom{2}}+p_{L2}^{\phantom{2}}}.$$

\begin{figure}
    \centering
    \begin{subfigure}[b]{0.48\textwidth}
        \includegraphics[width=\textwidth]{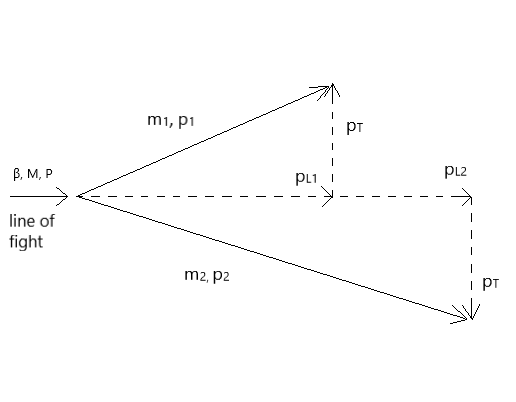}
        \caption{LAB}
        \label{fig:LAB}
    \end{subfigure}
    ~ 
    \begin{subfigure}[b]{0.49\textwidth}
        \includegraphics[width=\textwidth]{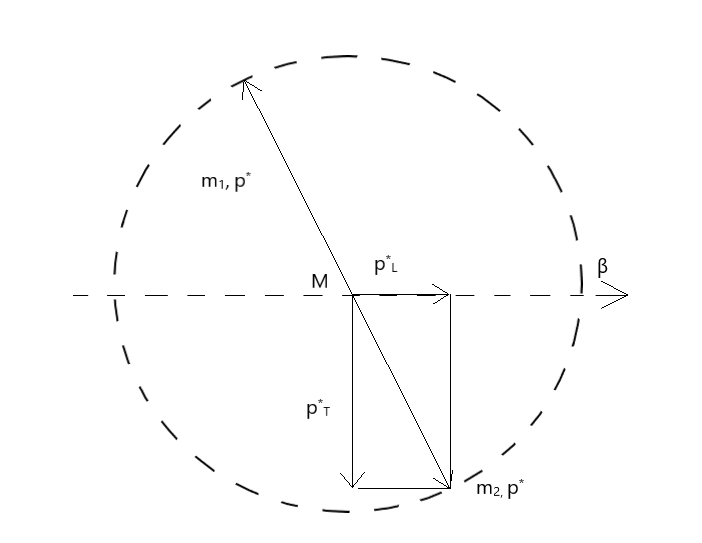}
        \caption{CM}
        \label{fig:CM}
    \end{subfigure}
    ~ 
    \caption{Kinematics of a two-body decay in (a) the laboratory (LAB) and (b) the center-of-mass (CM) reference frames.}\label{fig:refsistems}
\end{figure}

The longitudinal and transverse momenta, $p_L^{\phantom{2}}$ and $p_T^{\phantom{2}}$, are defined with respect to the direction of flight of the parent particle. In our case, the index 1 refers to the positively charged particle and the index 2 to the negatively charged particle. Figures~\ref{fig:LAB} and \ref{fig:CM} show the representations of the decay in the laboratory and center-of-mass reference frames, where the variables $p_L^{\phantom{2}}$ and $p_T^{\phantom{2}}$ are defined. The two particles have the same transverse momentum due to momentum conservation. In addition, due to the properties of Lorentz transformations, the transverse component of the boost remains invariant, thus $p_{T1}^{\phantom{2}} = p_{T2}^{\phantom{2}} = p_T^{\phantom{2}}$ and $p_T^{\phantom{2}}=p^*_T$.

\noindent 

\subsection{Parametric dependency of the ellipse}
Using the four-momentum conservation and the Lorentz transformations in the ultra-relativistic approximation ($\beta \rightarrow 1$), we can get the functional form of the decay:
\begin{equation}
\frac{(\alpha-\alpha_0)^2}{r_\alpha^2}+\frac{p_T^2}{{p^*}^2}=1,
\label{elipse}
\end{equation}

\noindent where all parameters are functions of the mass of the particles that participate in the decay: 
\begin{equation}
\alpha_0=\frac{m_1^2-m_2^2}{M^2}; \qquad r_\alpha=\frac{2p^*}{M}.
\label{defalpha}
\end{equation}
\begin{equation}
	{p^*}^2=\frac{1}{4M^2}(M^4+m_1^4+m_2^4-2M^2(m_1^2+m_2^2)-2m_1^2m_2^2).
\label{pcm}
\end{equation}

\noindent The masses of the parent and child particles can be obtained from a fit to the data with the above function. 
For the particular case of the $K_{S}^{0}\to \pi^{+}\pi^{-}$ decay, where the two final state products have the same mass $m\equiv m_{1}=m_{2}$, $\alpha_{0}$ goes to $0$ and Equation~(\ref{elipse}) simplifies to
\begin{equation}
\frac{\alpha^2}{r_\alpha^2}+\frac{p_T^2}{{p^*}^2}=1,
\label{elipse2}
\end{equation}
with 
\begin{equation}
	{p_T^2}=\frac{M^2}{4}(1 - \alpha^2)-m^{2}.
\label{pcm2}
\end{equation}

\noindent By fitting the above functions to data one can therefore determine the masses of all the particles involved in the decay, or alternatively, the momentum scale plus all the masses but one. In particular, for symmetric decays such as $K_{S}^0\rightarrow\pi^+\pi^-$, the position of the left and right vertices of the ellipse provides the mass of the parent in units of the child mass (no momentum scale involved), while the position of the top vertex provides a mass difference between the parent particle and the child mass in units of momentum as measured by the detector. It should be stressed that to obtain the momentum scale from the fit one needs one known particle mass even if many ellipses from different decays are simultaneously analyzed in an attempt to over-constrain the system of equations. This is because the momentum scale, when introduced in the formula, changes all the masses by the same amount -- as it is equivalent to a change of units -- hence the need of one external input to be able to gauge it.

\subsection{The Armenteros-Podolanski space in elliptical coordinates}\label{ellipticalArmenteros}
In this study, the masses are obtained by fitting the functional form of Equation~(\ref{elipse}) to data, which has the format
of 2D histograms. In the original Armenteros-Podolanski plane, most of the bins of these histograms are empty,
resulting in a waste of computing resources. To optimise the calculation, the data is transformed to elliptical coordinates, hereafter $r$ and $\phi$. We define a reference ellipse from known parameters (for example, the $K_S^0\rightarrow\pi^+\pi^-$ ellipse with world-averaged masses), and make a transformation such that this ellipse becomes a straight line ($r=1$). Then we classify the events in a 2D space with rectangular binning that covers such a line and its neighborhood. The transformation is done as follows:
\begin{eqnarray}
\alpha (r, \phi) = \tilde{\alpha}_0 + \tilde{r}_{\alpha} \, r \, \cos\phi, \nonumber\\
p_T^{\phantom{2}} (r, \phi) = \tilde{p}^* \, r \, \sin\phi.
\label{eliptical coordinatesalpha}
\end{eqnarray}
The constants $\tilde{\alpha}_0$, $\tilde{r}_{\alpha}$ and $\tilde{p}^*$ correspond to the values of $\alpha_0$, $r_{\alpha}$ and $p^*$ for the reference ellipse in which we want to focus our binning. The $r$ and $\phi$ coordinates are related to the Armenteros variables according to:
\begin{eqnarray}
r (\alpha, p_T^{\phantom{2}})= \sqrt{\frac{(\alpha - \tilde{\alpha}_0)^2}{\tilde{r}_\alpha ^2} + \frac{p_T^2}{\tilde{p}^{*2}}} , \nonumber \\
\phi (\alpha, p_T^{\phantom{2}}) = \arctan\left(\frac{p_T^{\phantom{2}} \, r_\alpha}{\tilde{p}^* \, (\alpha - \tilde{\alpha}_0)}\right).
\label{eliptical coordinatesphi}
\end{eqnarray}

\noindent The ellipses in the ($\alpha$, $p_T^{\phantom{2}}$) plane then transform to:
\begin{equation}
\frac{(\tilde{\alpha}_0 + \tilde{r}_{\alpha} \, r \, \cos\phi-\alpha_0)^2}{r_\alpha^2}+\frac{(\tilde{p}^* \, r \, \sin\phi)^2}{p^{*2}}=1
\label{elipsepasimetrica}
\end{equation}

\noindent This corresponds to a quadratic equation $A\, r^2\,+\, B \, r + \,C = 0$, whose positive solution is as follows:
\begin{eqnarray}
r = \frac{-B+\sqrt{B^2-4\, A\, C}}{2\, A}, \nonumber \\
A = \tilde{r}_\alpha^2 \, p^{*2} \, \cos^2\phi+r_\alpha^2\tilde{p}^{*2}\sin^2\phi, \nonumber \\
B = 2 \, \tilde{r}_\alpha \, p^{*2} \, \cos\phi (\tilde{\alpha}_0-\alpha_0) , \label{elipseeliptical} \\ 
C = p^{*2} \left( (\tilde{\alpha}_0-\alpha_0)^2 - r_\alpha^2 \right) . \nonumber
\end{eqnarray}

\noindent where $\tilde{\alpha}_0$, $\tilde{r}_{\alpha}$ and $\tilde{p}^*$ are
the constants of the transformation to elliptical coordinates. The parameters $\alpha_0$, $r_{\alpha}$ and $p^*$ specify the kinematics of a given $V^0$ decay and relate to the particle masses ($M$, $m_1$ and $m_2$) via Equations~(\ref{defalpha}) and (\ref{pcm}). For the particular case of the $K_{S}^{0}\to \pi^{+}\pi^{-}$ decay $m\equiv m_{1}=m_{2}$, Equation~(\ref{elipseeliptical}) simplifies to
\begin{eqnarray}
r = \sqrt{\frac{1}{\left(\frac{\tilde{r}_\alpha}{r_\alpha}\cos\phi\right)^2+\left(\frac{\tilde{p}^*}{p^*}\sin\phi\right)^2}}, \nonumber \\
r_\alpha = \frac{\sqrt{M^2-4 \, m^2}}{M}, \quad \tilde{r}_\alpha = \frac{\sqrt{\tilde{M}^2-4 \, \tilde{m}^2}}{\tilde{M}},   \\
p^* = \sqrt{M^2/4 - m^2}, \quad \tilde{p}^* = \sqrt{\tilde{M}^2/4 - \tilde{m}^2}.  \nonumber
\label{elipseeliptical2}
\end{eqnarray}

\begin{figure}
    \centering
    \includegraphics[width=\textwidth]{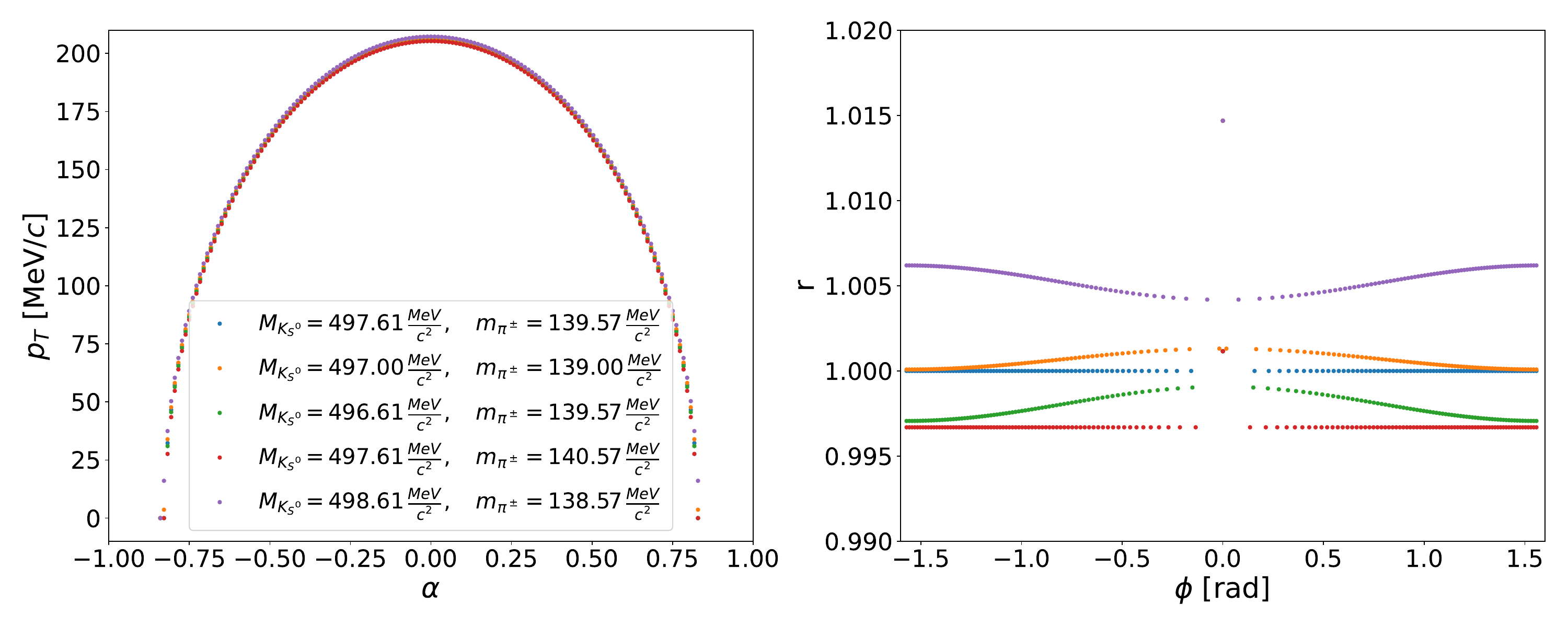}
   
\caption{Functional form of a set of symmetric decays ($m_1 = m_2 = m$) in the Armenteros plot (left) and in the plot with elliptical coordinates (right).}
\label{fig:Toys}    
\end{figure}

Figure~\ref{fig:Toys} shows the distribution  of $K_{S}^{0}\to \pi^{+}\pi^{-}$ decays generated in pseudoexperiments with similar mass parameters as the reference ellipse, in the Armenteros space and in the space with elliptical coordinates. While difficult to distinguish in the original Armenteros-Podolanski plane, the decays with similar but different values of the mass parameters can be easily separated by eye in the space with elliptical coordinates (for decays in the region of the Armenteros plot in which we are focusing). Shifts in the pion mass result in a vertical offset of the distribution but do not deform its shape, while shifts in the $K_{S}^{0}$ mass result in both a vertical offset and a deformation from a straight-line distribution.

Thanks to the fact that the functional shape of the decay is a horizontal line for the mass parameters that match the input parameters ($r_\alpha = \tilde{r}_\alpha,~\alpha_0 = \tilde{\alpha}_0,~p^{*2} = \tilde{p}^{*2} \Rightarrow r = 1$), in this space we can use a very fine rectangular binning in the fitting of the decay data.

\section{Fitting method through Theil index maximization}\label{cap:Theil}

The data is fitted with several hypotheses for the fit parameters, aiming at maximising the Theil index. This is a very useful technique for cases in which only one dominant component (in this case, the signal) is known, even if some outliers can be present due to background processes.

\subsection{Theil index}
The Theil index is a statistical magnitude within the framework of information theory that measures the redundancy of a system. It is widely used in economics to measure economic inequality in a population~\cite{Theil}. The Theil index is a scalar that measures, in entropy terms, how far is the system from the situation where all the parts of the system have the same value for a certain variable. Our system is the discrete Armenteros-Podolanski plot (a 2D-histogram) and the analyzed variable is the number of counts in each bin of the histogram ($x_i$).


If $N$ is the number of bins of the 2D-histogram, $x_i$ the number of counts in each bin $i$, and $\mu$ the overall average counts per bin, the normalized Theil index ($T$) is defined as:
\begin{equation}
T=\frac{1}{N\ln(N)}\sum_{i=1}^N\frac{x_i}{\mu}\ln(\frac{x_i}{\mu}),
\label{theileq}
\end{equation}
$$\mu=\frac{1}{N}\sum_{i=1}^Nx_i.$$

\noindent In this representation $T \in [0,1]$, where $T = 0$ means that the histogram is uniformly populated and $T = 1$ means that all the counts are in one bin of the histogram. To perform a fit with the Theil index as estimator, the fitting data is overlapped with pseudo-data generated from a hypothesis of the fit parameters. When the overlapped parameters reproduce the data, the Theil index will be maximized. This corresponds to a maximal inequality in the regions populated in the histogram as the fitting data and the pseudo-data fill the same bins of the histogram.

\subsection{Fit configuration}

Firstly, we generate a uniform binning of the Armenteros-Podolanski plot and we populate the 2D-histogram with the fitting data. Then we generate a large number of pseudo-data sets with mass hypotheses for the parent and child particles in given ranges. The array of pseudo-data points is generated uniformly in the $x$-axis of the plot, following the ellipse equation. Each set of pseudo-data points has the same number of events as the fitting data. Finally, for each mass hypothesis we sum the generated pseudo-data to the 2D-histogram of the fitting data and then we calculate the Theil index of each 2D-histogram.
The fit result will be the mass hypothesis that returns the maximum value of the Theil index. 

The fit parameters are $M, m_1, m_2$.
The fitting framework is based on graphics processing units (GPUs) and the fits are performed on an \texttt{Nvidia GTX 1080 Ti} GPU using the \texttt{Ipanema$-\beta$}~\cite{cuda_GPU} software.
The ranges for the mass hypotheses and the range of the plot considered for doing the 2D-histogram are reported in Table \ref{table:paramcalib}. The default number of hypotheses is 512\,000 and the default histogram binning is 150 x 150. Numerical precision is further improved by applying the GPU-based genetic algorithm~\cite{Chobanova:2017rkj} to the original population of 512\,000 mass hypotheses. The use of sampling techniques is motivated by the discrete values of the Theil index, which means that minimization algorithms based on gradients would experience convergence issues. 
If there is more than one mass hypothesis with the same maximum Theil index, we average them. 
The bias on the determination of the mass of a particle is defined as 
\begin{equation}
b = \frac{m-m_0}{m_0},
\label{mbias}    
\end{equation}
where $m$ is the result of the fit and $m_0$ is the known value of the particle mass. This quantity is a measurement of the momentum bias.

\subsection{Estimation of the statistical uncertainties}
The statistical uncertainties on the fit parameters are computed using the spread obtained in the fit result when applying the fit method to equally-sized subsets of the fitting data, each of them with its own set of pseudo-datasets. The different subsets are averaged, and the uncertainties on the mean mass values are applied as statistical uncertainties of the measurement.

The statistical uncertainty on $b$ is obtained through error propagation using  Equation~(\ref{mbias}),
$$\sigma(b) =  \frac{\sigma(m)}{m_0}.$$
In this study, we are using simulated data and the true value $m_0$ is exact, as given by Pythia 8.235~\cite{pythia}. In a scenario where real data is used, one needs to take into account the uncertainty on the value of $m_{0}$.
\subsection{Binning schemes}
\label{sec:binning_schemes}
In practice, we are searching for small deviations of the horizontal line $r=1$ in the elliptical Armenteros plane. Therefore, the binning in $r$ requires special attention to understand which range of sensitivities in the momentum scale can be achieved. If the range in $r$ is too narrow, most of the data will go out of bounds. On the contrary, if the number of bins is too small, the bin width will limit the momentum bias that can be resolved. We define an approximate range of validity of a binning scheme by:
\begin{equation}
    b^{max} \approx min(|r_{max}-1|,|r_{min}-1|)\\
\end{equation}

\noindent and

\begin{equation}
    b^{min} \approx (r_{max}- r_{min})/N_{r}
\end{equation}

\noindent where $N_r$ is the number of bins in $r$.
The binning schemes used in this study are listed in Tab~\ref{table:paramcalib}. Each binning scheme covers a range of roughly two orders of magnitude of the potential bias of the momentum scale.

\begin{table}[]
\centering
\caption{Binning schemes. The name of each scheme follows the notation ``BN'', where $N\approx\log_{10}{b^{min}}$ taking into account that the used number of bins is $N_r = 150$.}
\begin{tabular}{c|c|c|c|c}
                      & $\textbf{r}$& $\bm{\phi}$  & $\sim b^{max}$ & $\sim b^{min}$\\ \hline

\textbf{B-8} & $[0.9999994$,$\, 1.0000006]$     & $[-\pi/2$,$ \, \pi/2]$ & $6\times 10^{-7}$& $8\times 10^{-9}$\\ 
\textbf{B-7} & $[0.999994$,$\, 1.000006]$     & $[-\pi/2$,$ \, \pi/2]$ & $6\times 10^{-6}$& $8\times 10^{-8}$\\
\textbf{B-6} & $[0.99994$,$\, 1.00006]$     & $[-\pi/2$,$ \, \pi/2]$      & $6\times 10^{-5}$ & $8\times 10^{-7}$               \\
\textbf{B-5} & $[0.9994$,$\, 1.0006]$     & $[-\pi/2$,$ \, \pi/2]$ & $6\times 10^{-4}$ & $8\times 10^{-6}$          
                \\
\textbf{B-4} & $[0.994$,$\, 1.006]$     & $[-\pi/2$,$ \, \pi/2]$ & $6\times 10^{-3}$ & $8\times 10^{-5}$                     
\end{tabular}
\label{table:paramcalib}
\end{table}

\section{Simulated data sample}
\begin{figure}
    \centering
    \includegraphics[width=\textwidth]{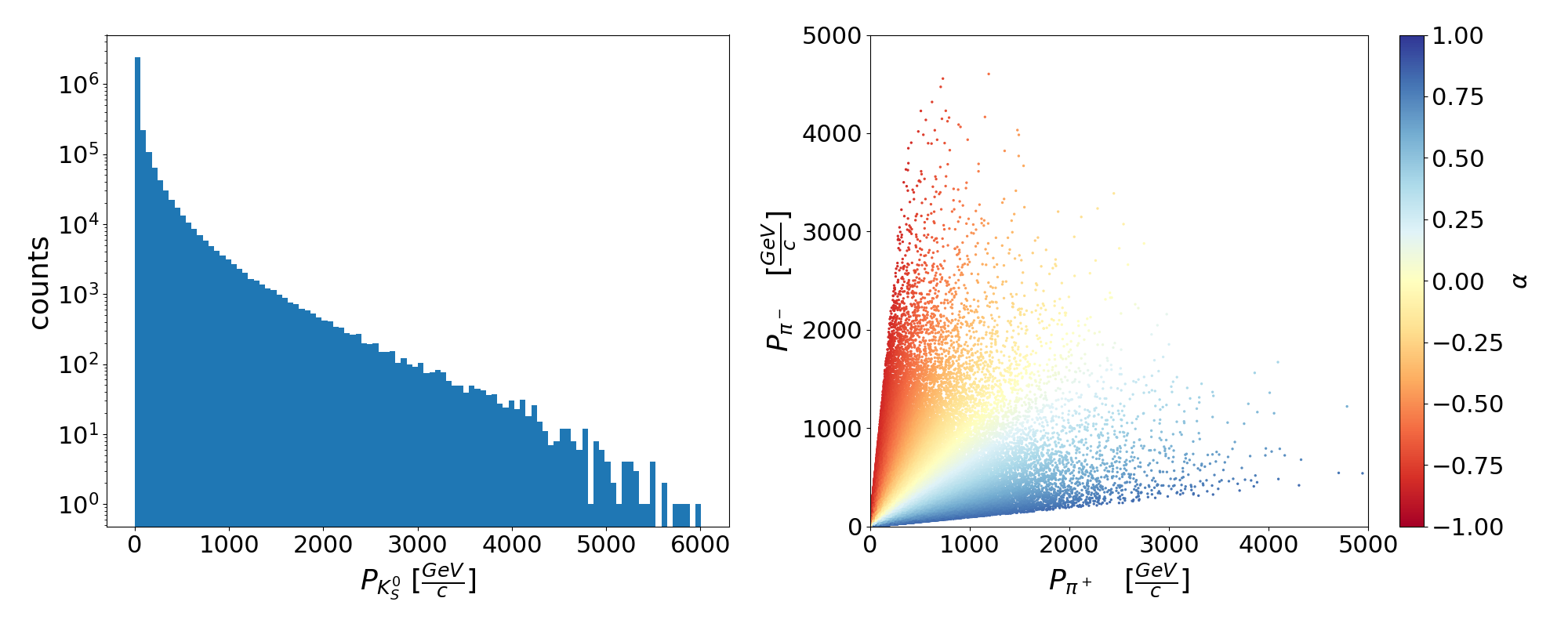} 
\caption{Momentum distribution of simulated $K_S^0\to \pi^{+}\pi^{-}$ decays originating from $\sqrt{s}=13$ TeV $pp$ collisions for (left) $K_S^0$ and (right) pions. Also, the dependency of the pion momenta on the Armenteros variable $\alpha$ is shown.}
\label{fig:data-distributions}    
\end{figure}

The data used to test the method are a sample of $3$ million $K_S^0\to \pi^{+}\pi^{-}$ decays simulated with Pythia~\cite{pythia}, in $\sqrt{s}=13$ TeV $pp$ collisions. The momentum distribution of the $K_S^0$ and pions in data is shown in Fig.~\ref{fig:data-distributions}. These distributions exhibit a long tail, corresponding to large momentum values. The $K_S^0$ in these long tails agree better with the ultra-relativistic approximation ($\beta \rightarrow 1$) used to build the ellipse in the Armenteros space and lead to a better fit performance. Conversely, the bulk of the data does not match this approximation and appears as noise in the fit. The data points that are in a better agreement with this approximation will be those that contribute to the signal.

\section{Fit Performance}

In this section we evaluate the performance of the fit program by analysing unbiased simulated samples and search for residual biases and their statistical precision. Two types of residual biases are expected in this simplified study.
\begin{itemize}
    \item Biases due to the finite bin size. As discussed in Section~\ref{sec:binning_schemes}, each binning scheme is sensitive down to $b_{min}$ level. Residual biases of these size are therefore expected.
    \item Biases due to the $\beta=1$ approximation. For simplicity we are using an ultra-relativistic approach in our fit formula, while the analyzed samples are generated with Pythia at finite energies.
\end{itemize}

\subsection{Fit to monochromatic simulated data samples}

\begin{figure}[!h]
	\includegraphics[width=9 cm]{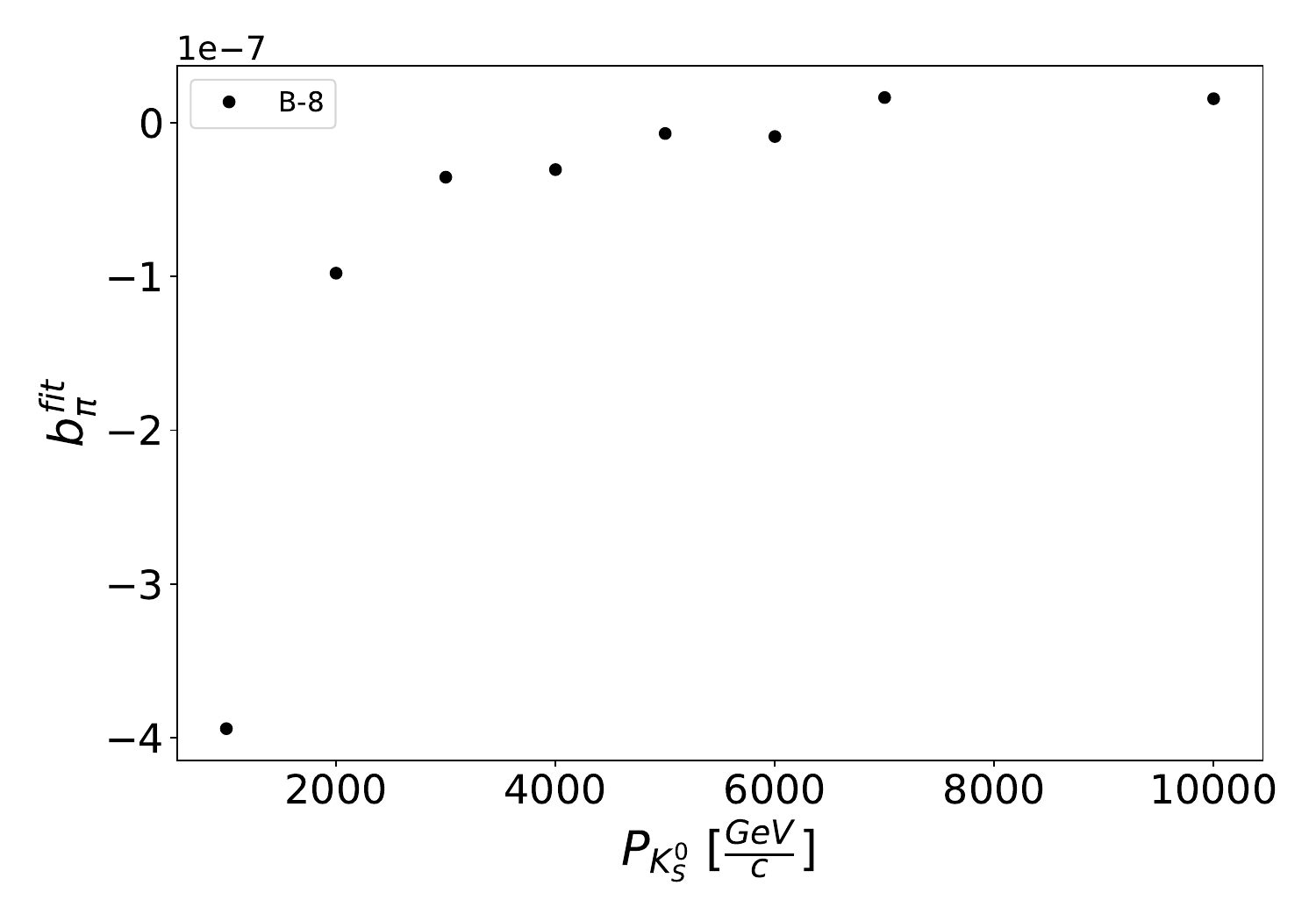}
	\centering
	\caption{Dependence of $b_{\pi}$ on the $K_S^0$ momentum of simulated data.}
\label{fig:bias-boost}
\end{figure}

The effect of the $\beta$-value of the parent in the fit is shown in Fig.~\ref{fig:bias-boost}, where we simulate many decays with a monochromatic momentum of the $K_S^0$, then fit the corresponding ellipse in the Armenteros space with elliptical coordinates using binning scheme {\bf B-8}. It can be seen that the bias is larger at low kaon momenta, while it asymptotically converges to a value near the $b_{min}$ of {\bf B-8} at larger momenta.
This effect could be corrected by using a fitting formula that is valid for any $\beta$-value and does not yield in general a semi-ellipse in this space. In that way, our fit would take into account all the possible $\beta$-values for our data, or by imposing tight requirements in the momentum. But as shown in the next subsection the $\beta=1$ approach is sufficiently sensitive for the purpose of this study.

\subsection{Fit to unbiased simulated data samples} \label{cap:fit_to_armenteros_ellip}

In Fig.~\ref{fig:bias-boost} we have seen the bias as a function of the momentum of monochromatic $K_S^0$. In samples generated with Pythia at LHC energies the kaons have many posible momentum values, and the residual bias due to the $\beta=1$ approach is not obvious. It has to be stressed that the Theil index maximization will search for a region with the expected shape, hence the presence of a large enough number of ultra-relativistic kaons may be enough for the residual bias to be negligible. In this section we evaluate the effect of the $\beta=1$ approach on the sample of 3 million decays generated with Pythia in $p-p$ collisions at 13 TeV.
We fit the generated data sample in the elliptical coordinates using the ultra-relativistic approximation ($\beta \rightarrow 1$) and without any cut on the kaon momentum values.

 We use binning schemes {\bf B-6} and {\bf B-7}.
 The region that is important in the fit is the region with very high $\beta$ values and high density of counts, although we have a lot of data points with different $\beta$ values. Our method is successful in this situation because the Theil index is blind to the data points that deviate far from the signal region where the density of data points is very low.

\begin{figure}
    \centering
    \includegraphics[width=\textwidth]{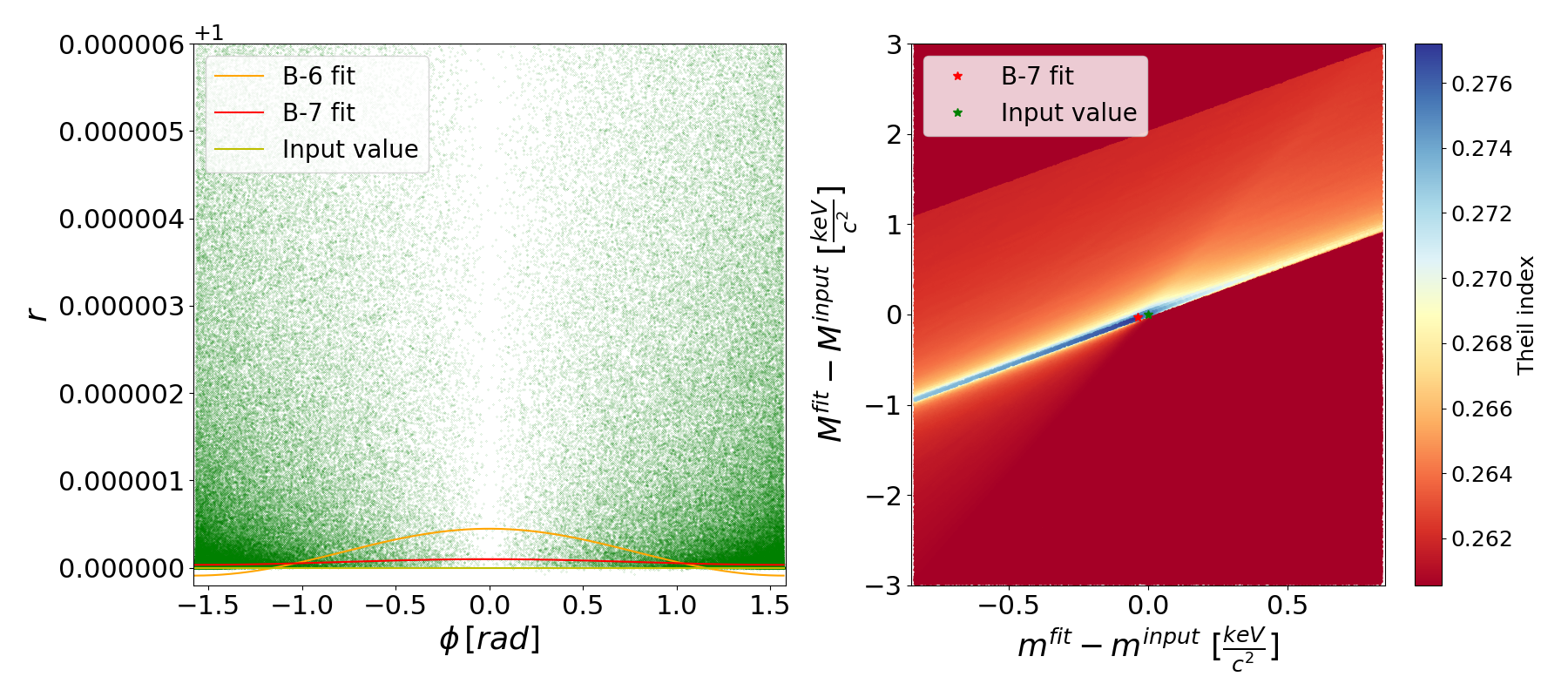}
   
\caption{Left: Projection of the fit result in the elliptical Armenteros-Podolanski plane. In the plot are represented the fit results using {\bf B-6} and {\bf B-7} binning schemes and the equation $r=1$ that is obtained with the input value of the mass parameters. Right: Theil index as a function of the mass hypothesis values using {\bf B-7} binning scheme on the fit to all the data.}
\label{fig:armenterosplot}   
\end{figure}

The uncertainty is estimated using 25 equally-sized datasets.
The fit result is shown in Fig.~\ref{fig:armenterosplot} and in Table~\ref{table:fit12res}. The left-hand plot shows the fit results in the Armenteros space with elliptical coordinates.The right-hand plot shows the mass parameter space with a color scale that represents the value of the Theil index for each mass hypothesis done in the calculation with binning scheme {\bf B-7}. This representation helps to highlight the region with the highest values of the Theil index (``signal region'').

\begin{table}[]
\centering
\caption{Fit results obtained on section~\ref{cap:fit_to_armenteros_ellip}. The uncertainties on the fit result appears in parentheses.}
\begin{tabular}{l|l|l|l}
                    & \textbf{$\textbf{M} \, [\frac{\rm MeV}{c^2}]$} & \textbf{$\textbf{m} \, [\frac{\rm MeV}{c^2}]$} & $b_{\pi}$ \\ \hline
\textbf{Input value}  & $497.61$                          & $139.57$                            & $-$               \\
\textbf{\textbf{B-6} fit}       & $497.609733(32)$                      & $139.569789(21)$                           & $-1.52(15) \, 10^{-6}$         \\
\textbf{\textbf{B-7} fit}       & $497.6099677(30)$                      & $139.5699608(25)$                           & $-2.81(18) \, 10^{-7}$
\end{tabular}
\label{table:fit12res}
\end{table}

\begin{figure}
    \centering
    \includegraphics[width=\textwidth]{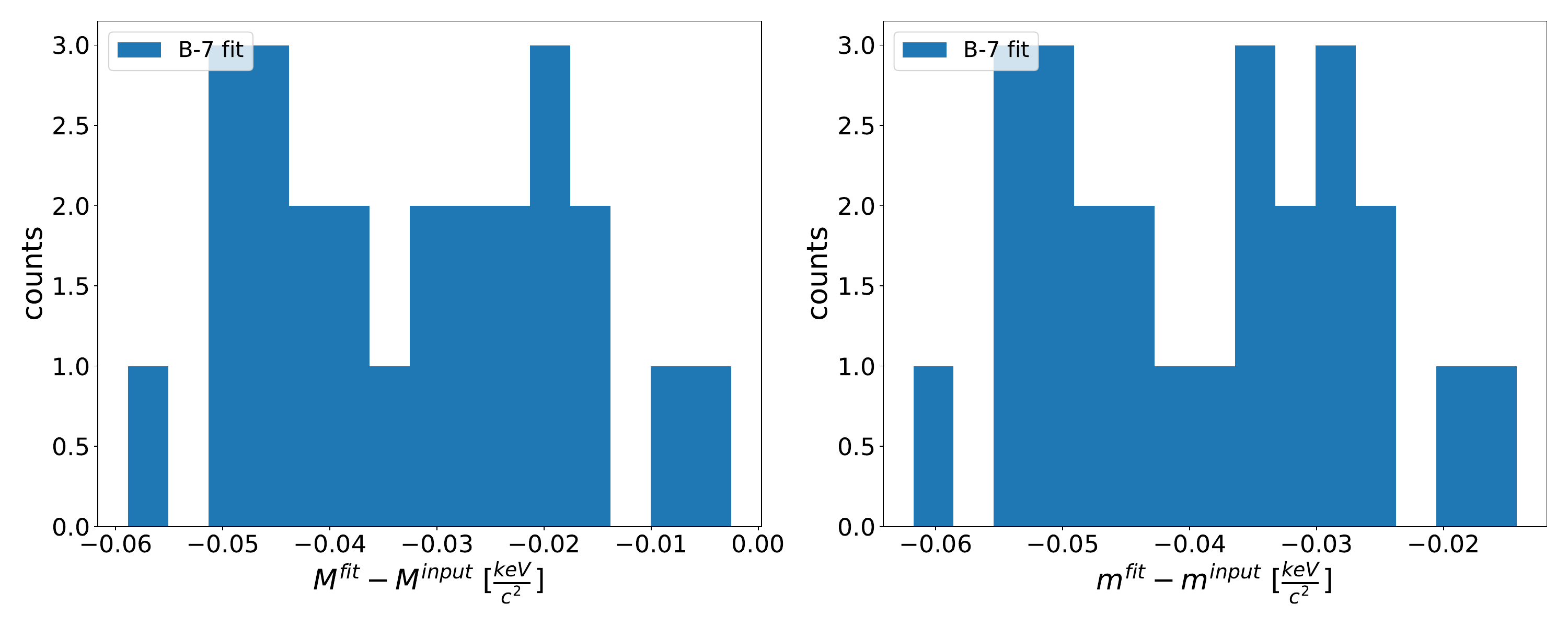}
   
\caption{Distribution of (left) $M_{K_S^0}$ and (right) $m_{\pi^\pm}$ outcomes in 25 different subsamples using {\bf B-7} binning scheme. The values obtained on the fit are subtracted by the input value.}
\label{fig:elipticalincert}    
\end{figure}

The histograms used to assign the statistical uncertainties on $m$ are shown in Fig.~\ref{fig:elipticalincert}. They are obtained from fits to subsets of the data in the space with elliptical coordinates. We note that the statistical uncertainty is dominated by the number of pseudo-datasets used to calculate the Theil index, rather than due to the sample size of the simulated data in Pythia.

\section{Performance of the calibration of the momentum scale} \label{cap:Calibracion}

We now assess the sensitivity of the method to perform the calibration of the momentum scale of a particle physics detector.

To perform the simulation of the calibration of the momentum scale, we start from the momenta of the child particles, supposing these are known exactly. Then, we simulate a simplified deviation by multiplying the values of all the momenta of particles by the same value of the momentum scale for each component of the momentum, $p_x,~ p_y,~ p_z$. 
The study is repeated for several biases on the momentum scale.
The last step is to construct the Armenteros space with these new values of the children's momenta and then perform the fits. The goal is to reproduce the deviation applied to the momentum scale accurately.

We now evaluate the performance of the method in the calibration of the momentum scale. We introduce a bias in the momentum scale by multiplying the momentum values by a certain factor, and the goal is to be sensitive to these deviations in the fit result. We multiply all the momentum values of the children of the simulated decays by the momentum scale factor, $S_{p}$. If $p'$ is the true momentum value and $p$ the momentum value with the applied bias in the momentum scale then:
\begin{equation}
p_{ij} = S_{p} p'_{ij} ; \quad i=\pi^+, \, \pi^-; \quad j = x, \, y, \, z.
\label{MS_equation}
\end{equation}

At this point we can define the bias in the momentum as:
\begin{equation}
b_{p} = \frac{p - p'}{p'} = S_{p} - 1 
\label{p_bias}
\end{equation}

\begin{figure}[!h]
	\includegraphics[width=11 cm]{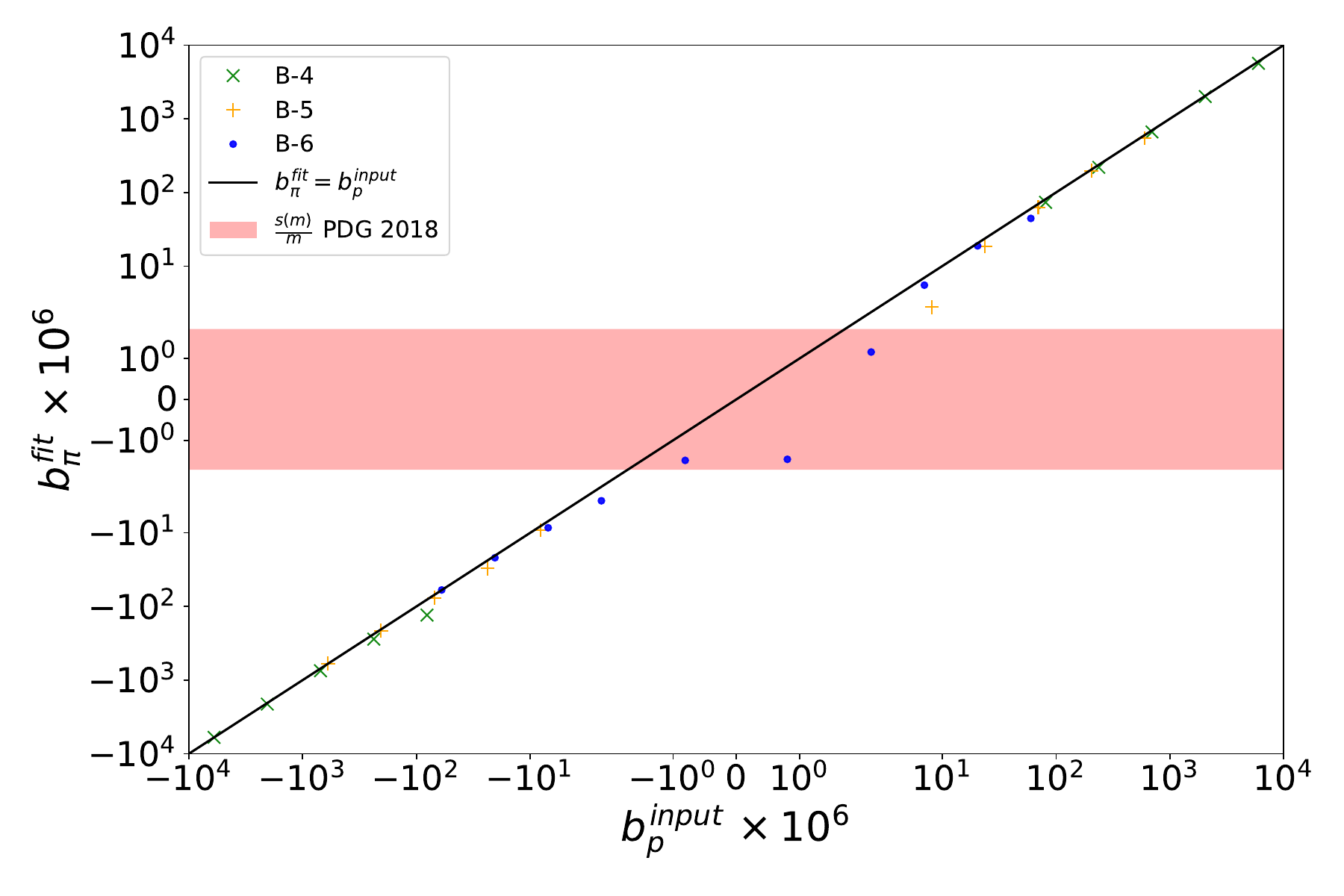}
	\centering
	\caption{Performance of the calibration of the momentum scale of a detector. The $x$-axis shows the applied $b_p$ multiplied by a factor $10^6$ to improve the visualization of the plot. The $y$-axis corresponds to the bias of the child's mass parameter obtained in the fit multiplied by a factor $10^6$. Both axes are in semi-logarithmic scale. The relative uncertainty in the knowledge of the pion mass is represented by a red rectangle.}
\label{fig:MOMENTUMSCALE2}
\end{figure}

 The fit result used is the mean in the fit to 3 million events divided into 25 data sets.
Figure~\ref{fig:MOMENTUMSCALE2} shows the result in the performance of the calibration. With \textbf{B-6} binning scheme  (shown in Table~\ref{table:paramcalib}), the limitation in the calibration is the precision with which the mass of the pion meson is known. According to Ref.~\cite{PDG2018} the mass value of the pion is $m_{\pi^\pm} = $139.57061(24) MeV$/c^2$ which corresponds to a relative uncertainty of $s(m)/m \approx 10^{-6}$. This limits the precision that we can achieve in the knowledge of the momentum scale, which can be seen in the plot to be around $10^{-5}-10^{-6}$. 
\noindent Higher precision can be further achieved by using better known masses, such as protons from $\Lambda^0$ decays or muons from charmonia or bottomonia states.

In addition, it is also important to remark that the calibration lines in the plot are not symmetric (i.e. do not pass though the centre of coordinates), which is due to the existence of points that do not match the ultra-relativistic approximation ($\beta \rightarrow 1$), and hence the $b_m$ is underestimated as long as we stick to the ultra-relativistic description.

\subsection{Fit for the momentum scale} \label{cap:FitMomentumScale}
Up until now, we fitted the masses in the Armenteros-Podolanski plot and then we obtained the bias in the momentum scale through the bias in the fit of the pion mass parameter. Now, we are going to present an alternative method where we can estimate the momentum scale directly from the fit. Hence, we have to include the $S_p$ parameter defined previously in the fit function. Considering the case where the bias on the momentum scale is the same for the three directions of the momentum, we can solve for the $S_p$ factor from the $p_T^{\phantom{2}}$ Armenteros variable, and the $\alpha$ variable is not affected by the $S_p$ factor as it cancels out in the ratio of momenta.
As the parameter that we use for the calibration in Figure~\ref{fig:MOMENTUMSCALE2} is the mass of the charged pion ($m_{\pi^{\pm}}$), we fix this parameter. Hence, to obtain the momentum scale, we fit the for $S_p$ factor as well as the (in principle, unbiased) mass of the parent particle ($M_{K_S^0}$), which becomes a nuisance parameter for our purpose.

Following the notation in Equation~\ref{MS_equation},
$$p_T^{\phantom{2}} = p'_T \, S_p \, ,\qquad \alpha = \alpha' .$$
By introducing these parameters in the ellipse equation (Equation~(\ref{elipse})), we obtain the fitting function, which is then transformed into elliptical coordinates. In the case of $K_S^0 \rightarrow \pi^{+} \pi^{-}$, the function can be written as:
\begin{eqnarray}
r = \sqrt{\frac{1}{\left(\frac{\tilde{r}_\alpha}{r_\alpha}\cos\phi\right)^2+\left(\frac{\tilde{p}^*}{\bm{S_p} \, p^*}\sin\phi\right)^2}}, \nonumber \\
r_\alpha = \frac{\sqrt{M^2-4 \, m^2}}{M}, \quad \tilde{r}_\alpha = \frac{\sqrt{\tilde{M}^2-4 \, \tilde{m}^2}}{\tilde{M}},   \\
p^* = \sqrt{M^2/4 - m^2}, \quad \tilde{p}^* = \sqrt{\tilde{M}^2/4 - \tilde{m}^2}.  \nonumber
\label{MS_fit_equation}
\end{eqnarray}

The fits of this plot were done to 3 million data points divided in 25 subsets for each momentum scale.

\begin{figure}[!h]
	\includegraphics[width=\textwidth]{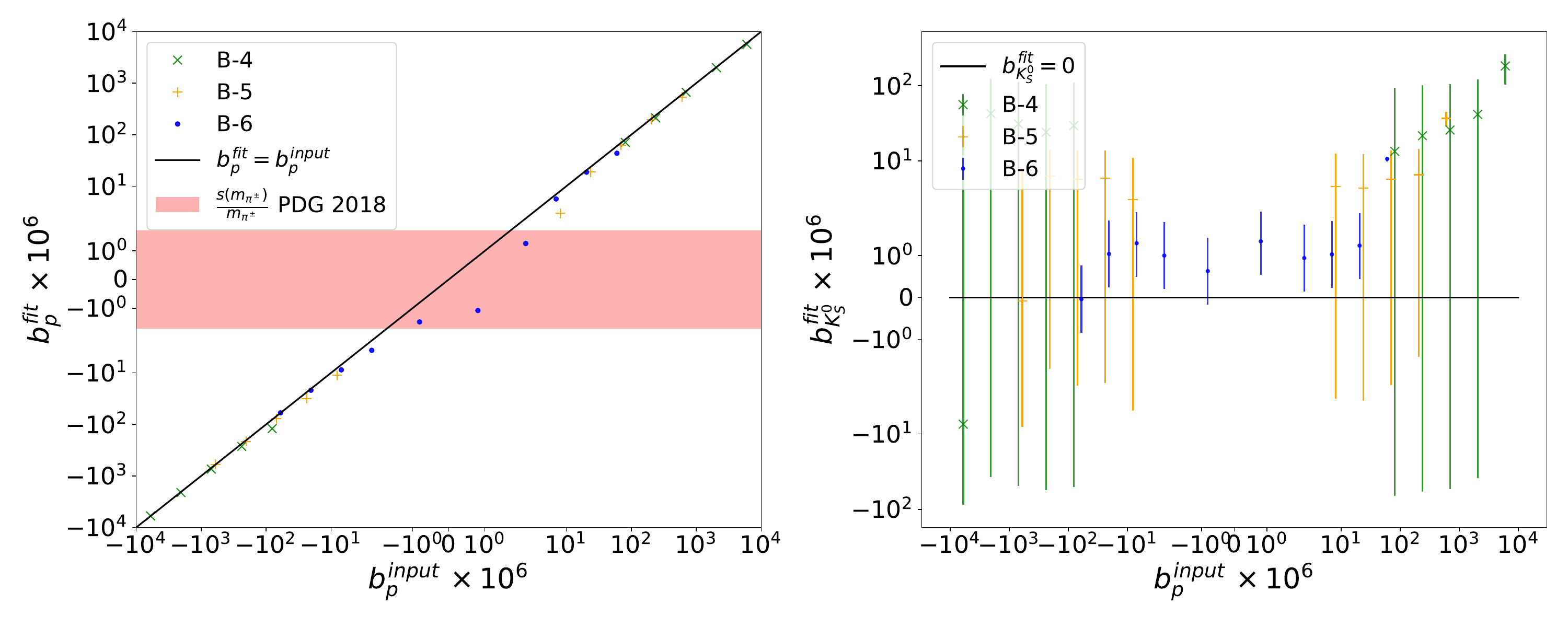}
	\centering
	\caption{Performance of the calibration of the momentum scale of a detector. The $x$-axis shows the applied momentum bias in semi-logaritmic scale and multiplied by $10^6$. The $y$-axis shows the $S_p$ parameter of the fit subtracting 1 (equal to $b_p$ on the fit) and multiplied by $10^6$ to improve visualization. The relative uncertainty in the knowledge of the pion mass is represented by a red rectangle. In the plot of the right is represented the obtained biases of the mass of the parent particle. A systematic uncertainty on the knowing of the mass parameter is assigned, the value os this uncertainty is equal to $b^{min}$ for each binning scheme. The binning schemes used are shown in Table~\ref{table:paramcalib}}
\label{fig:MS_fit}
\end{figure}

As shown in Fig.~\ref{fig:MS_fit}, the method behaves well in the whole range of the fit except for the central region ($\approx 10^{-6}$) range. This is because the presence of many data points that do not match the ultra-relativistic approximation.
In Fig.~\ref{fig:MS_fitmonocromatic}, we use a monochromatic data-sample, i.e. a sample where all the parent particles has the same momentum ($P_{K_S^0} = 5 \, TeV/c$). In this plot can be seen that the deviation in the central region becomes smaller.  
\textbf{
\begin{figure}[!h]
	\includegraphics[width=11 cm]{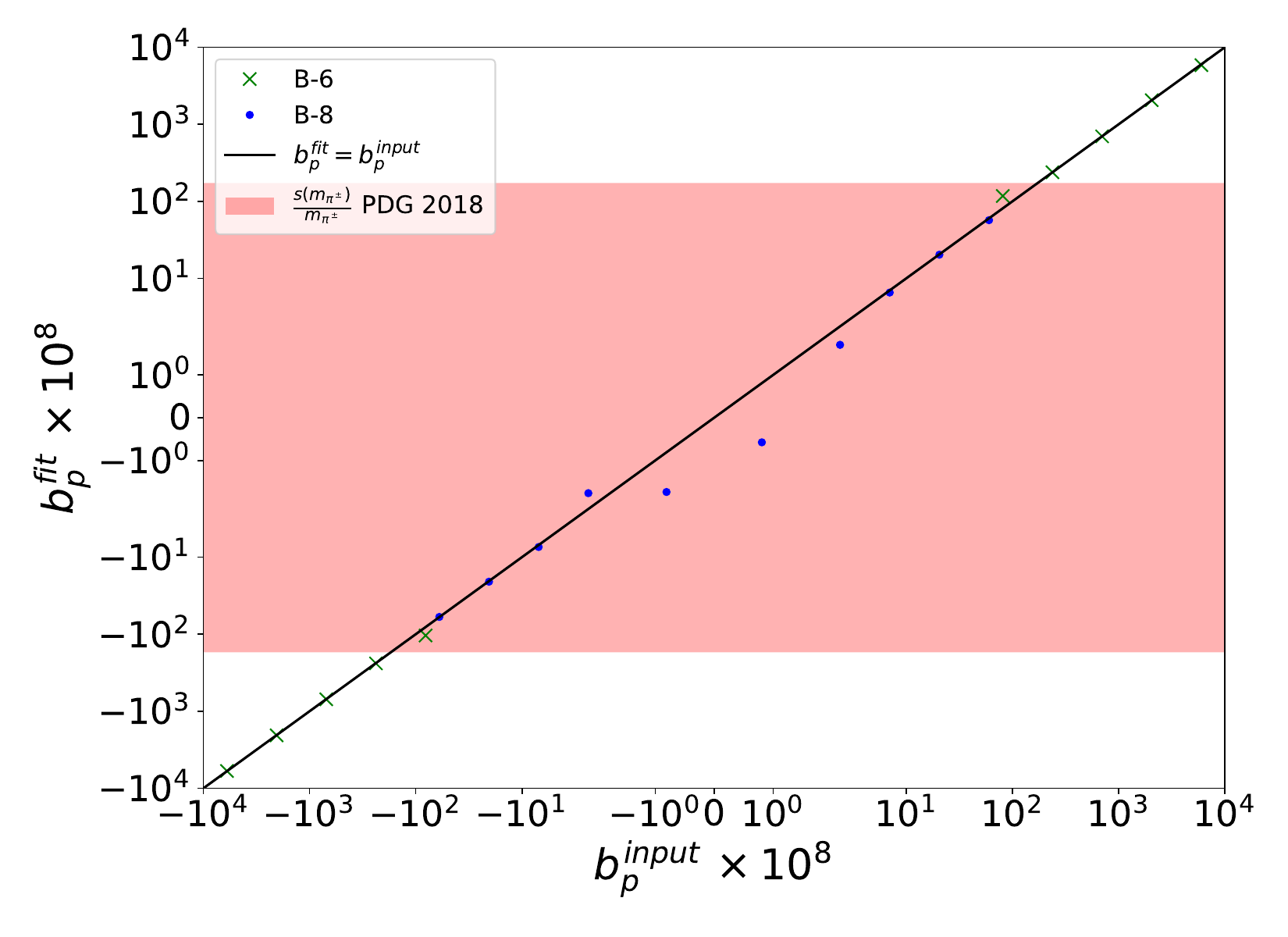}
	\centering
	\caption{Performance of the calibration of the momentum scale where the data-sample is monochromatic, being the momentum of all the parent particles $P_{K_S^0} = 5 \, TeV/c$. The $x$-axis shows the applied momentum bias in semi-logarithmic scale and multiplied by $10^8$. The $y$-axis shows the bias obtained on the fit multiplied by $10^8$ to improve visualization. The relative uncertainty in the knowledge of the pion mass is represented by a red rectangle. The used binning schemes are shown in Table~\ref{table:paramcalib}}
\label{fig:MS_fitmonocromatic}
\end{figure}}

\begin{figure}[!h]
	\includegraphics[width=0.48\textwidth]{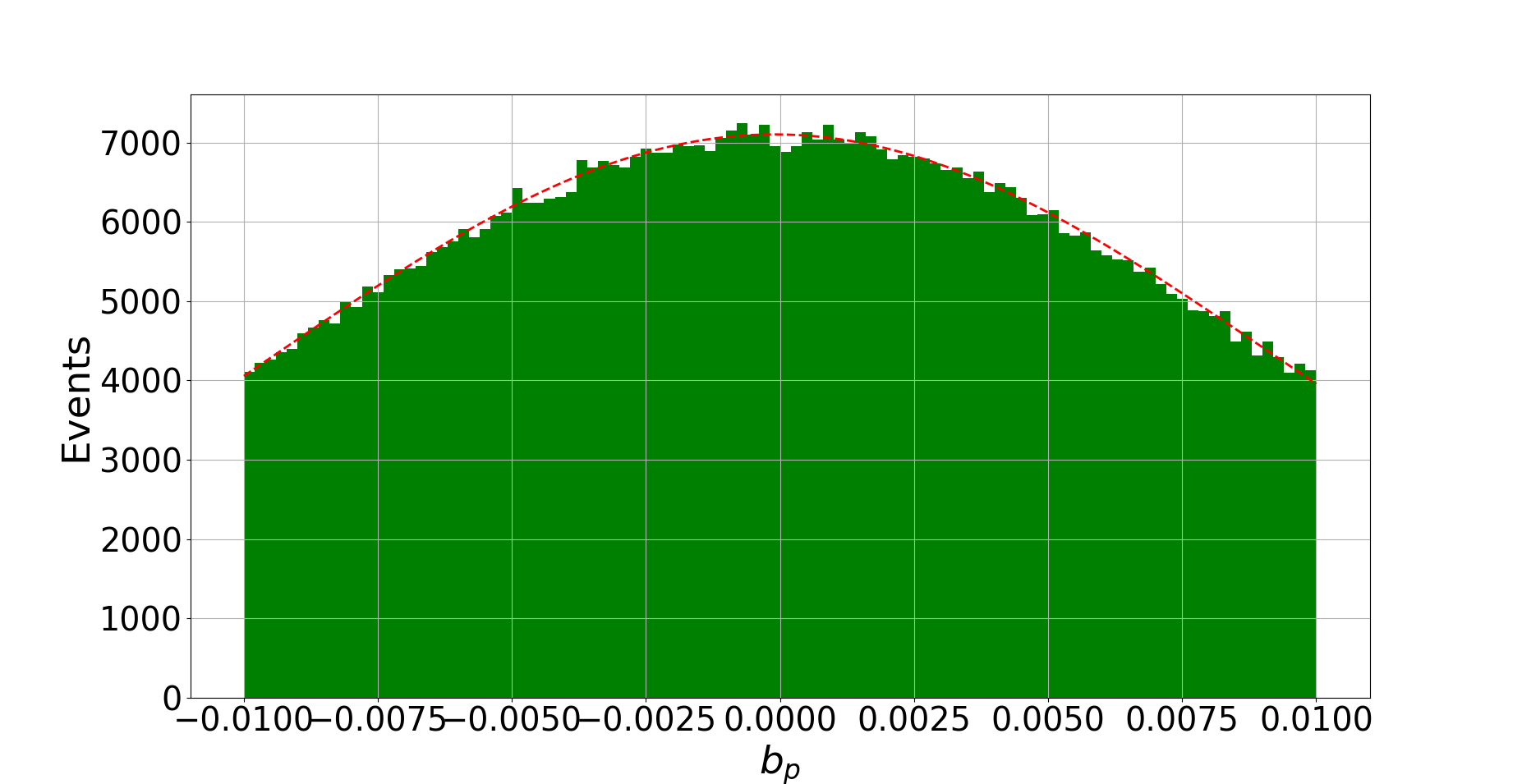}
	\includegraphics[width=0.48\textwidth]{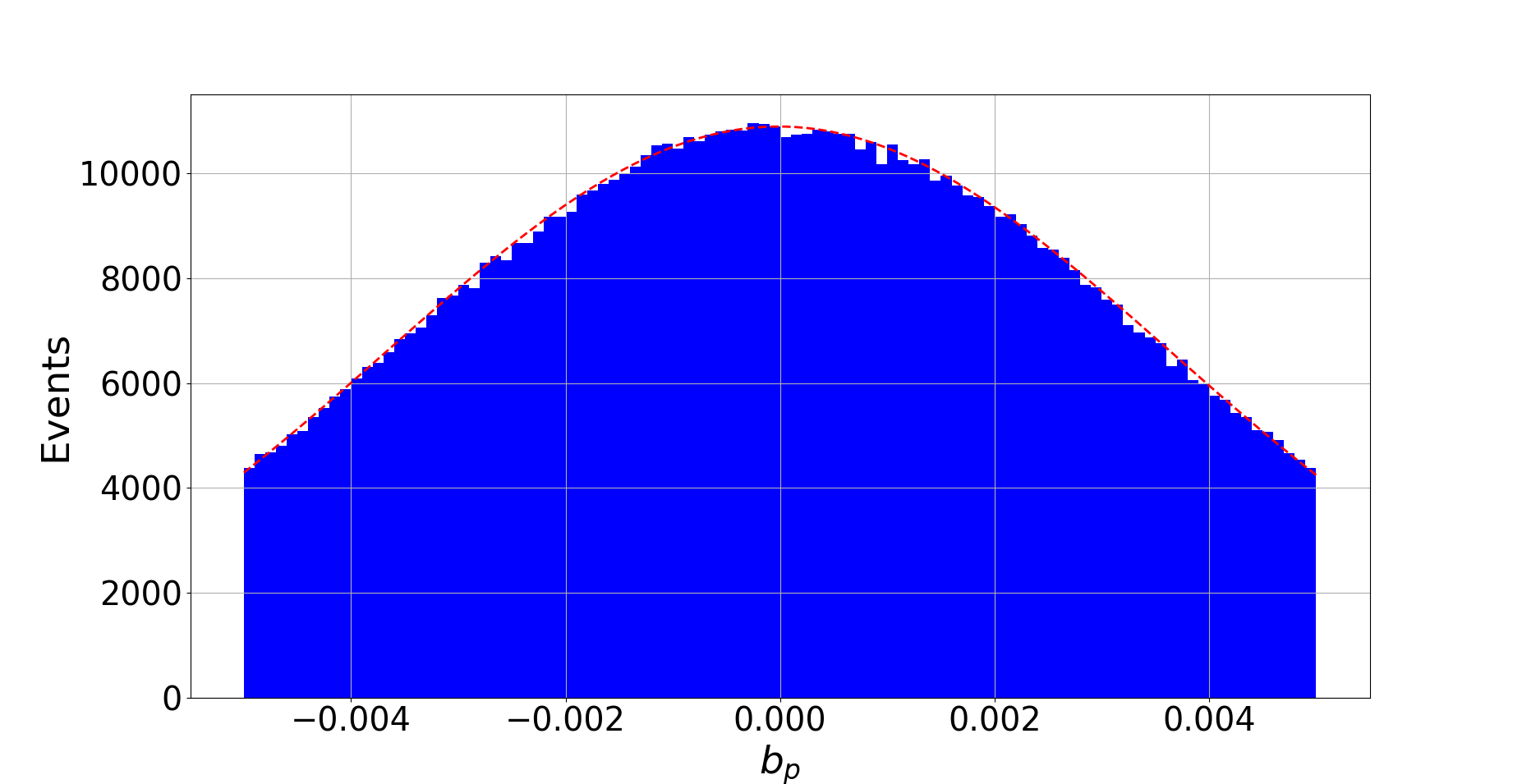}
	\centering
	\caption{Distribution of $b_p$ in $K_S^0\rightarrow\pi^+\pi^-$ (left, green) and  $J/\psi\rightarrow\mu^+\mu^-$ (right, blue) decays simulated through an LHCb-like detector using the software in ~\cite{Chobanova:2020vmx}. The value of $S_p$  used to obtain $b_p$ in a per-event basis is calculated following Eq. ~\ref{eq:per_event_nobeta}.  The red dashed line shows a Gaussian distribution fitted to the simulated data }
\label{fig:per_event}
\end{figure}

On a real experiment, the statistical uncertainty would be mainly determined by the momentum resolution and the sample size. With our current setup we can't generate, model, and fit a detailed simulation of the detector effects into the Armenteros plane. However, we can still assess the effect of the momentum resolution. In the hypothetical case in which both the mass of the parent particle and the mass of the child particle are well known, the information of the 2D plane can be condensed in a single variable:
\begin{equation}
    S_p = \frac{\sqrt{M^2(1-\alpha^2) - 4 \, m^2}}{2 \, p_T^{\phantom{2}}} 
    \label{eq:per_event}
\end{equation}

\noindent or, more generally without the ultrarelativistic approximation:
\begin{eqnarray}
S_p = \sqrt{\frac{-B+\sqrt{B^2-4\, A\, C}}{2\, A}}, \nonumber \\
A = \frac{4 \, p_T^2}{M^2} , \nonumber \\
B = \frac{4 \, p_T^2}{P^2} + \alpha^2 - 4 \, (\frac{M^2}{4} - m^2) \, \frac{1}{M^2} , \label{eq:per_event_nobeta} \\ 
C = - 4 \, (\frac{M^2}{4} - m^2) \, \frac{1}{P^2} ; \nonumber
\end{eqnarray}

\noindent which corresponds to a per-event estimate of the momentum scale. For perfectly reconstructed events and perfectly known masses, the distribution of S would be a delta function peaking at the value of the momentum scale. For realistic events, which are subject to momentum resolution, the distribution of $S_p$ is smeared, as illustrated in Fig. ~\ref{fig:per_event}. Such a distribution has a mean and a standard deviation, from which the uncertainty of the mean can be obtained as the standard deviation divided by the square root of the number of events. For one million $K_S^0\rightarrow\pi^+\pi^-$ decays simulated through an LHCb-like detector ~\cite{Chobanova:2020vmx}, and approximating the distribution near the peak to a Gaussian distribution (so that we don't have to make distinctions between mean and mode), we obtain an uncertainty of about $9\times 10^{-6}$ (see Fig. ~\ref{fig:per_event}. Hence, one would need about 100 million of $K_S^0\rightarrow\pi^+\pi^-$ decays to get a statistical uncertainty on the momentum scale bellow $10^{-6}$ , ten billions to get bellow $10^{-7}$, and one trillion to get bellow the $10^{-8}$ level, which is already much more precise than the world knowledge of the pion mass. Similarly, one would need about 10 million $J/\psi\rightarrow\mu^+\mu^-$ decays to get a statistical uncertainty on the momentum scale bellow $10^{-6}$, one billion to get bellow $10^{-7}$, and 100 billions to get bellow the $10^{-8}$ level. The needed number of decays is smaller for the $J/\psi$ case compared to the $K_S^0$ case because the mass resolution is about three times better in terms of the absolute mass of the decaying particle.

\section{Discussion and conclusions}

In this work, we studied a method for improving the precision of the momentum scale of a particle physics detector. We presented a method, in which we fit the masses in the Armenteros-Podolanski space with a transformation of variables that allows us to improve the precision in the binning of the space. This is a key feature since in this space the decay of interest appears as an horizontal line which allows us to do a fine rectangular binning around the data points of interest of the decay. We illustrated the method using $K_S^0\rightarrow\pi^+\pi^-$ decays.

The motivation for using the Armenteros space rather than the traditional fits to well-known resonances is that this allows us to use the mass of the decay products for performing the calibration, such as that of the pion. 
We perform the study in an ideal case. To adapt the method to a more realistic situation, one could, for instance, generate the pseudo-experiments using a fast simulation that takes into account experimental effects, such as acceptance, momentum resolution, internal alignments, energy loss, path through the magnet, and radiation of soft bremsstrahlung photons in the $V^0$ decays. The ultimate precision of the method can be further by fitting for the mass of the proton in the $\Lambda^0 \rightarrow p\pi^{-}$ decay, which is known to a precision better than $10^{-8}$, or by fitting for the mass of the muon -known at the $10^{-8}$ level- in $J/\psi\rightarrow\mu\mu$ decays. Interestingly, one can also look at the mass of the same child particle (for example, the muon), at different mass peaks (e.g., charmonia and bottmonia) to check for kinematic dependency of the momentum bias as well as for self-consistency of the method. Of course, the final accuracy of  the obtained result will depend on how systematic effects related to detector description are handled.
The implementation on GPUs facilitates the speedy generation of a large number of mass hypotheses with in a fine binning. 
Moreover, using this approach, a single experiment can determine the masses of many observed particles using its own data (and understanding of its detector effects) and relying only on a single external input (which could be chosen to be, for example, the well-known proton or muon masses). The method can also be used to search for beyond-SM quasi-stable charged particles.

\section{Acknowledgements}
We would like to thank M. Needham for useful feedback in our draft. This work has received financial support from Xunta de Galicia (Centro singular de investigaci\'on de Galicia accreditation 2019-2022), by European Union ERDF, and by  the “Mar\'ia  de Maeztu”  Units  of  Excellence program  MDM-2016-0692  and  the Spanish Research State Agency. In particular: the work of X.C.V. is supported by MINECO (Spain) through the Ram\'{o}n y Cajal program RYC-2016-20073 and by XuntaGAL under the ED431F 2018/01 project; the work of V.C. is supported by MCINN (Spain) through the Juan de la Cierva-incorporaci\'{o}n program with grant IJCI-2017-32371.  V.V.G. is supported by the grant ERC-CoG-724777 RECEPT. V.V.G., D.M.S., M.L.M., and J.M. acknowledge the hospitality of the Petnica Science Centre during the development of the concept behind this paper in 08/2015. M.L.M acknowledges support from NWO (Netherlands).

\clearpage

\bibliographystyle{unsrt}
\bibliography{main}

\begin{thebibliography}{10}

\bibitem{armenteros}
J.~Podolanski and R.~Armenteros.
\newblock {III. Analysis of V-events}.
\newblock {\em The London, Edinburgh, and Dublin Philosophical Magazine and
  Journal of Science}, 45(360):13--30, 1954.

\bibitem{detector_performance}
Roel Aaij et~al.
\newblock {LHCb Detector Performance}.
\newblock {\em Int. J. Mod. Phys.}, A30(07):1530022, 2015.

\bibitem{Aaij:2011ep}
R.~Aaij et~al.
\newblock {Measurement of $b$-hadron masses}.
\newblock {\em Phys. Lett.}, B708:241--248, 2012.

\bibitem{pythia}
Torbjorn Sjostrand, Stephen Mrenna, and Peter~Z. Skands.
\newblock {A Brief Introduction to PYTHIA 8.1}.
\newblock {\em Comput. Phys. Commun.}, 178:852--867, 2008.

\bibitem{pion_mass}
M.~Trassinelli et~al.
\newblock {Measurement of the charged pion mass using X-ray spectroscopy of
  exotic atoms}.
\newblock {\em Phys. Lett.}, B759:583--588, 2016.

\bibitem{PDG2018}
M.~Tanabashi et~al.
\newblock {\href{http://pdg.lbl.gov/}{Review of particle physics}}.
\newblock {\em Phys. Rev.}, D98:030001, 2018.

\bibitem{Theil}
Henri Theil.
\newblock {\em The Information Approach to Demand Analysis}, pages 627--651.
\newblock Springer Netherlands, Dordrecht, 1992.

\bibitem{cuda_GPU}
D.~Martínez Santos, P.~Álvarez Cartelle, M.~Borsato, V.~G. Chobanova,
  J.~García Pardiñas, M.~Lucio Martínez, and M.~Ramos~Pernas.
\newblock {Ipanema$-\beta$ : tools and examples for HEP analysis on GPU}.
\newblock arXiv:1706.01420, 2017.

\bibitem{Chobanova:2017rkj}
Veronika Chobanova, Giancarlo D'Ambrosio, Teppei Kitahara, Miriam
  Lucio~Martinez, Diego Martinez~Santos, Isabel~Suarez Fernandez, and Kei
  Yamamoto.
\newblock {Probing SUSY effects in $K_S^0\rightarrow\mu^+\mu^-$}.
\newblock {\em JHEP}, 05:024, 2018.

\bibitem{Chobanova:2020vmx}
Veronika Chobanova, Diego~Mart\'\i{}nez Santos, Claire Prouve, and Marcos
  Romero~Lamas.
\newblock {Fast simulation of a forward detector at 50 and 100 TeV
  proton-proton colliders}.
\newblock arXiv:2012.02692, 12 2020.

\end{thebibliography}

\end{document}